\documentclass[prl,twocolumn,showpacs,preprintnumbers,amsmath,amssymb,floatfix]{revtex4}
\usepackage{hyperref}
\usepackage{epsfig}
\newcommand{\rem}[1]{}

\begin{document}

\title{Dynamical chiral symmetry breaking in sliding nanotubes}

\author{X.~H. Zhang$^{1,2}$, G.~E. Santoro$^{1,2,3}$, U. Tartaglino$^{2,1}$, and E. Tosatti$^{1,2,3}$}
\affiliation{$^1$ International School for Advanced Studies (SISSA), Via Beirut 2-4, I-34014 Trieste, Italy}
\affiliation{$^2$ CNR-INFM Democritos National Simulation Center, Via Beirut 2-4, I-34014 Trieste, Italy}
\affiliation{$^3$ International Centre for Theoretical Physics (ICTP), P.O.Box 586, I-34014 Trieste, Italy}

\date{\today}

\begin{abstract}
We discovered in simulations of sliding coaxial nanotubes an unanticipated example 
of dynamical symmetry breaking taking place at the nanoscale. While both nanotubes 
are perfectly left-right symmetric and nonchiral, a nonzero angular momentum of 
phonon origin appears spontaneously at a series of critical sliding velocities, 
in correspondence with large peaks of the sliding friction. The non-linear equations 
governing this phenomenon resemble the rotational instability of a forced string.
However, several new elements, exquisitely ``nano'' appear here, with the crucial 
involvement of Umklapp and of sliding nanofriction.
\end{abstract}

\maketitle


A popular high-school physics demo is a clamped oscillating rope or guitar string.
While being imparted a strictly planar vibration at one end, the string
initially vibrates, as expected, within the plane. Above a certain 
amplitude however the plane of vibration spontaneously and unexpectedly 
begins to turn around, right or left with equal probability \cite{elliott.ja:1982}.
Since all along the string's Lagrangian (including the external forcing)
remains completely left-right symmetric (i.e., nonchiral) this is a prototype
example of what may be called {\it dynamical chirality breaking} taking place in
the macroscopic world. Dynamical spontaneous symmetry breaking of chiral symmetry 
is actually rife in nature, including examples such as the Taylor-Couette instability 
in hydrodynamics \cite{Guyon}, the spontaneous chirality of vibrating cilia 
in biology\cite{lenz.p:2003} and many other macroscopic scale examples.

We discovered, in simulations of the frictional sliding of carbon nanotubes, a
nanoscale example of dynamical chiral symmetry breaking in particular in two
coaxial nanotubes, one forced to slide inside 
the other. While both nanotubes are perfectly left-right symmetric and nonchiral, 
angular momentum strikingly jumps to nonzero values in correspondence to
some sliding velocities, coincident with large phonon-related peaks of the sliding friction.
The theory of this phenomenon yields non-linear equations that differ from the 
string problem by newer elements that are exquisitely nano, now involving 
Umklapp processes and sliding nanofriction.


\begin{figure}[!tb]
  \centering
    \epsfig{file=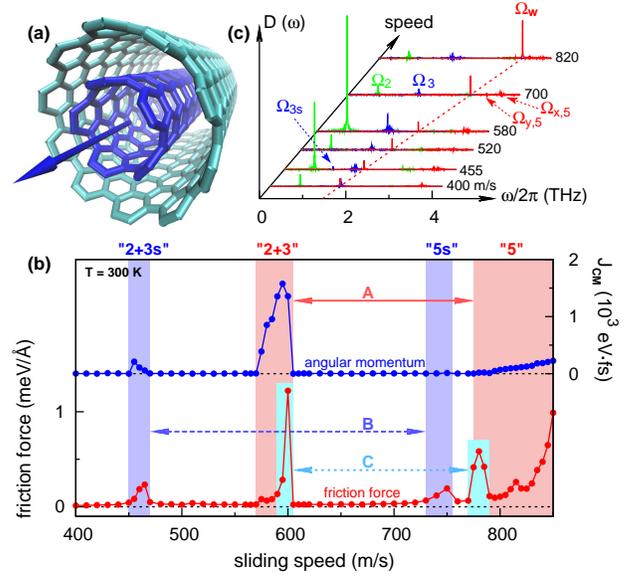,width=8.3cm,angle=0,clip=}
  \caption{\label{fig.force.and.momentum}
    (a) Coaxial sliding of (5,5)@(10,10) 
    nanotubes. (b) Sliding friction 
    force per inner tube atom and angular momentum $J_{\rm CM}$ from simulations 
    versus speed $v$ at $T=300$ K.  Note the frictional peaks and the threshold near 780 m/s.  
    The nonzero angular momentum signals nanoscale dynamical chiral symmetry breaking 
    in correspondence with the threshold, and with peaks at 570 m/s and 450 m/s.
    A, B, and C designate related resonance regions, described in text.
    (c) Outer tube radial motion Fourier spectra with $n=2$ (green), 3 (blue), and 5 (red) symmetry, 
    showing resonant enhancement in correspondence with peaks and threshold.
    }
\end{figure}

We conducted classical molecular dynamics simulations of an inner, infinite and 
termination-free, (5,5) armchair single-wall carbon nanotube sliding at speed $v$ 
inside a coaxial (10,10) nanotube, as in Fig.~\ref{fig.force.and.momentum}a. 
Standard empirical potentials were used for the intratube \cite{brenner.dw:1990} and 
intertube \cite{kolmogorov.an:2005} interactions. 
%
%
Details of simulation are given in Ref.~\onlinecite{zhang_ecoss}.
Temperature was controlled, usually at $T$ = 300 K, by an algorithm designed 
to preserve angular momentum \cite{soddemann.t:2003}.
As noted earlier by Tangney {\it et al.} \cite{tangney.p:2006}, the tube-tube
sliding friction is not hydrodynamical, in fact not even monotonic with $v$,
but develops sharp peaks and onsets at selected speeds. We found sharp
frictional peaks near $v$ = 450 m/s, 570 m/s, 720 m/s, and an important
threshold onset near 780 m/s, shown in Fig.~\ref{fig.force.and.momentum}b.
These peaks are known to generally arise out of parametric excitation of outer nanotube 
``breathing'' phonon modes, classified by an angular momentum index $n$ 
(for tangential quantization around the tube axis). 
Our peak positions differ from earlier ones \cite{tangney.p:2006} due to our 
lack of tube terminations (implying mode uniformity along the tube),  
and also to our different inter-tube potentials.  
As noted earlier \cite{zhang_ecoss}, the non-monotonic 
$F$--$v$ characteristics implies a ``negative differential friction'', whereby an increasing
applied force $F$ yields an innner-tube velocity that grows by jumps and plateaus, rather than smoothly.

The surprise comes from analysing, at the frictional peaks, the 
two parts of angular momentum $J$ around the tube axis $y$: its center-of-mass 
(rigid body rotation with angular velocity $\omega$) and shape-rotation (``pseudo-rotational'') 
parts $J=J_{\rm CM}+J_{\rm pseudo}$,  with $J_{\rm CM} = \sum_i m_i [ \vec{r}_i \times 
(\vec{\omega}\times\vec{r}_i) ]_y$, and $J_{\rm pseudo} = \sum_i m_i ( \vec{r}_i \times \dot\vec{r}_i )_y$.
Generally zero at generic $v$ due to lack of nanotube chirality, 
$J_{\rm pseudo} = - J_{\rm CM} = 0$, $J_{\rm pseudo}$ jumps to nonzero values at 
the frictional peaks and past the threshold, where $J_{\rm pseudo} = - J_{\rm CM} \ne 0$ 
(total $J$ is clearly conserved), see Fig.~\ref{fig.force.and.momentum}b.
%
Simulations at lower temperatures (not shown) reveal other smaller frictional peaks, 
but the main findings of the present paper, including chirality breaking, are 
still present, and indeed even stronger.

\begin{figure}[!tb]
  \centering
    \epsfig{file=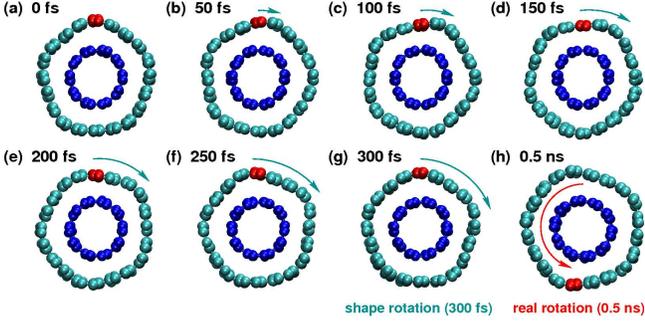,width=8.5cm,angle=0,clip=}
  \caption{\label{fig.pseudorotation}
    Tube cross-sections, showing a shape-rotation (pseudo-rotation) and an overall
    rotation (marked by red atoms) for $v=820$ m/s 
    (washboard period
    $a/v = 300$ fs). (a)--(g) Snapshots at intervals of $50$ fs comprise one period
    of pseudo-rotation, where the pentagon shape rotates clockwise. 
    (h) An overall counter 
    rotation of nearly $\pi$ in 0.5 ns.}
\end{figure}

Cross-section snapshots at intervals of $50$ fs, at $v=820$ m/s of 
Fig.~\ref{fig.pseudorotation} show a large pentagonal distortion of the outer 
tube, corresponding to an $n=5$ breathing mode. A vertex of the pentagon moves clockwise
(or counter-clockwise with 50-50 probability in different simulations)
with angular velocity $\Omega_w/5$, where $\Omega_w=2\pi v/a$ is the ``washboard'' 
frequency associated with the sliding speed $v$ 
and $a=2.46$ {\AA} is the intertube potential periodicity.
The pentagon rotation signals a non-vanishing 
$J_{\rm pseudo}>0$. Since the total $J=0$, the center-of-mass acquires an equal and opposite 
counter-rotation $J_{\rm CM}=-J_{\rm pseudo}<0$, as demonstrated by 
the red atom marker in Fig.~\ref{fig.pseudorotation}.

\begin{figure}[!tb]
  \centering
    \epsfig{file=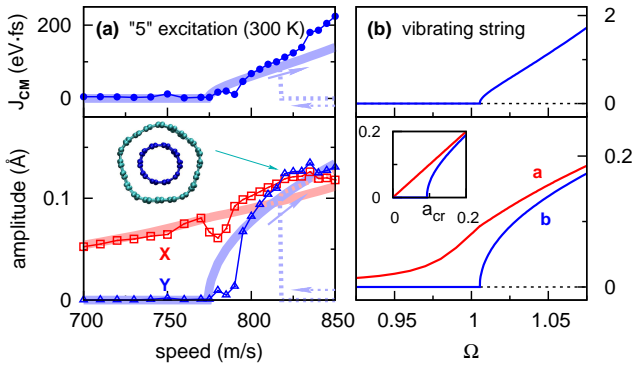,width=8.5cm,clip=}
  \caption{\label{fig.amplitudes}
    (a) 
    Simulation results (dots) in the threshold region 
    $v>780$ m/s,
    compared with the model (thick solid and dotted lines) of Eqs.~(\ref{n=5:eqn}). 
    $X,Y = |u_{z,5,+} \pm u_{z,5,-}^*|$ are
    amplitudes of 
    $X(t)$ and $Y(t)$ defined in the text.
    $Y$ becomes nonzero together with nonzero 
    $J_{\rm CM}=-J_{\rm pseudo}$ above the critical threshold speed. 
    The hysteresis found in the model is absent in the simulation.
    (b) Forced vibrating string from Eqs.~(\ref{string:eqn}), 
    with $\omega_{k,T}=1$, $K=5$, $F=0.002$, and $M=100$.
    $a,b$ are the $X,Y$-mode amplitudes.
    The critical frequency $\Omega_{\rm cr} = 1.0054$ and the threshold
    $X$-mode amplitude $a_{\rm cr} = 0.093$ (inset).}
\end{figure}

{\it Theory} - The mechanism underlying symmetry breaking is the tube non-linearity, 
reminiscent of the string instability \cite{elliott.ja:1982}.
The non-linear motion of a string forced to vibrate
along a transverse direction $x$ at frequency $\Omega$ is described by 
\begin{eqnarray} \label{string:eqn}
\ddot X &=& - [\omega_{k,T}^2 + K(|X|^2+|Y|^2)] X + F \cos(\Omega t) \nonumber \\
\ddot Y &=& - [\omega_{k,T}^2 + K(|X|^2+|Y|^2)] Y, 
\end{eqnarray}
where $\omega_{k,T}=k\sqrt{\mu/M}$ is the bare transverse frequency with wavevector 
$k$ ($\mu$ is the Lam\'e constant and $M$ the mass density), $X, Y$ are two orthogonal 
linear polarizations, and $F$ is the forcing amplitude. 
The nonlinearity $K>0$ accounts for the increase of the transverse frequency 
at large amplitudes, where the larger overall elongation leads to an effective 
string tension increase. 
Due to $K>0$ there is a critical frequency $\Omega_{\rm cr}=\sqrt{\omega_{k,T}^2 + (KF^2/16)^{1/3}}$
beyond which the string, although forced along $x$, spontaneously develops both $X$ and $Y$ modes, 
an elliptic polarization resulting from a purely linear forcing, with a spontaneous dynamical chirality breaking.
The critical amplitude of the $X$-mode beyond which the elliptical polarization sets in is $(2F/K)^{1/3}$
(this will be demonstrated in Fig.~\ref{fig.amplitudes}b below). 
In the nanotube case, the sliding inner tube excites phonon modes of the outer one,
uniformly along the tube 
in our termination-free case. 
The outer (10,10) nanotube has, among others, doubly-degenerate 
modes with angular momentum $n=\pm 2$, $\pm 3$, $\pm 4$ and $\pm 5$, etc. 
Let $u_x(\theta)$ and $u_z(\theta)$ be displacements in the $x$ (tangential) and $z$ 
(radial) directions at angular position $\theta$ on the outer tube circumference, 
and $u_{n,x,\pm}$ and $u_{n,z,\pm}$ their respective $n$-mode amplitudes, 
travelling clockwise ($+$) and counter-clockwise ($-$). 
Suzuura and Ando (SA) \cite{suzuura.h:2002} described these modes at the quadratic level 
in a continuum model. Third- and fourth-order energy terms in $u$ mix the different $n$-modes in all
possible ways compatible with conservation of angular momentum. The main nonlinear terms 
$\propto \int dx (\frac{\partial u_x}{\partial x} + \frac{u_z}{R})
                 (\frac{\partial u_z}{\partial x} - \frac{u_x}{R})^2$ to third order, and
$\propto \int dx (\frac{\partial u_z}{\partial x} - \frac{u_x}{R})^4$ to fourth order
turn out to be as crucial as in the string.


{\it ``$\mathit{n=5}$'' excitation} - The 10-chain atomic structure of the sliding (5,5) inner 
tube stimulates at all speeds $v$ the outer tube with a deformation corresponding to a 
linearly polarized, $k_y =0$, $n=5$ mode with a washboard frequency $\Omega_w = 2\pi v/a$. 
Away from resonance, $\Omega_w < \omega_5$, the excitation amplitude 
of the $n=5$ outer tube mode is small, and linear. 
Near and above resonance, however, the amplitude grows, and nonlinearities become dominant.  
Starting from the SA model \cite{suzuura.h:2002} including non-linearities, 
after some approximations (mainly $u_{x,n,\pm} \approx (i/n) u_{z,n,\pm}$ as 
appropriate from linear eigenvector analysis, neglecting small mode-mixing 
third-order terms while keeping large fourth-order contributions which generate 
terms such as $|u_{z,5,+/-}|^2 u_{z,5,+}$), we arrive at equations for the 
$n=\pm 5$ phonon amplitudes of the suggestive form
\begin{eqnarray} \label{n=5:eqn}
  \ddot{X} &=& - [\omega_5^2 + Q + (P/4) (|X|^2 + |Y|^2)] X + 4F \cos(\Omega_w t) \nonumber \\
  \ddot{Y} &=& - [\omega_5^2 - Q + (P/4) (|X|^2 + |Y|^2)] Y,
\end{eqnarray}
where the two variables $X(t) = (u_{z,5,+} + u_{z,5,-}^*) e^{i\Omega_w t} + c.c.$ 
and $Y(t) = i (u_{z,5,+} - u_{z,5,-}^*) e^{i\Omega_w t} + c.c.$ parameterize the radial 
displacement $u_{z,5}(\theta,t)$ in terms of two orthogonal standing waves, 
$u_{z,5}(\theta,t) = X(t) \cos(5\theta) + Y(t) \sin(5\theta)$, 
and the fourth-order term $P$ again shifts upwards the frequency as the amplitude increases. 
%
%
The important difference between Eqs.~(\ref{n=5:eqn}) and those of the string is the ``nano'' $Q$-term, 
representing an {\it Umklapp} process coupling two outer tube modes with $n=\pm 5$ through a 
``reciprocal lattice vector'' of the inner tube, with its 10 carbon double-chains.
The Umklapp term splits the $X/Y$ frequencies to approximately $\sqrt{\omega_5^2 \pm Q}$. 
A static double-tube phonon calculation confirms split frequencies $\omega_x=3.35$ 
and $\omega_y=3.15$ THz which measure the Umklapp strength.

In nanotube sliding, unlike a guitar string that can be pinched soft or hard,
the only controllable parameter is the speed $v$. Nevertheless, due to the similar 
fourth-order non-linear effects, the physics of nanotube sliding, 
in the region with $v>780$ m/s (shaded and labelled ``$5$'' in Fig.~\ref{fig.force.and.momentum}b), 
resembles that of the string, see Fig.~\ref{fig.amplitudes}a. 
Near the 780 m/s threshold, $\Omega_w\approx 3.17$ THz, just above 
$\omega_y=3.15$ THz, a spontaneous symmetry breaking occurs, the $Y$-amplitude 
growing from zero, and the center-of-mass angular momentum with it. 
In the approximations considered so far, this should correspond to a pure ``$n=5$'' excitation; 
and indeed this is close to reality, as shown in the Fourier spectrum of 
Fig.~\ref{fig.force.and.momentum}c, where at $v=820$ m/s the most important 
peak appears at 
the washboard frequency $\Omega_w$. Note that the effective $n=5$ mode frequency
is dragged along by the washboard, growing with $v$ and with the friction magnitude.
When the speed grows larger than 830 m/s, simulations show the additional excitation of $n=2$ 
or $n=3$ modes, due to third-order non-linearites not accounted for in Eqs.~(\ref{n=5:eqn}). 
Here the dynamics, still chiral, becomes more complex than that of the pure ``$n=5$'' mode. 
In Fig.~\ref{fig.amplitudes}a solutions of Eqs.~(\ref{n=5:eqn}) are shown as shaded 
thick lines, with parameters $P=(1728/50) \alpha/MR^4$, $Q$, and $F$
which were fit to the bulk modulus of the nanotube ($\alpha=15000$ {\AA}$^2/{\rm ps}^2$), 
the intrinsic phonon frequencies and the splitting for $n=5$ ($Q=50.5/{\rm ps}^2$), 
as well as the energy corrugation ($F=5$ \AA$^2/{\rm ps}^2$) of the inter-tube potential. 
The model agrees fairly well with hard simulation results, including a crossing of
$X$ and $Y$ amplitudes at around $v=815$ m/s, which only occurs due to Umklapp. 
A feature predicted by the model is a hysteretic behaviour with respect to increasing 
and decreasing $v$. The hysteresis is not seen in simulations both at $300$ K and at $50$ K. 
We suspect that frictional Joule heating could be sufficient to remove it.
Also, our model being mean-field in character, it does not allow
for fluctuations. The simulations of course do, and indeed occasional reversals
of the sign of $J_{\rm CM}$ are observed when $|J_{\rm CM}|$ is very small; 
these reversals are suppressed, most likely by inertia, when  $|J_{\rm CM}|$ is large.  
 
\begin{figure}[!tb]
  \centering
    \epsfig{file=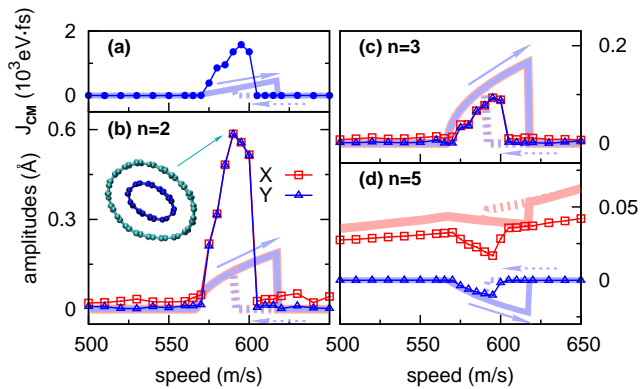,width=8.5cm,clip=}
  \caption{\label{fig.amplitudes.2+3}
    Simulation results (dots) and comparison with the model (with a velocity backshift of 78 m/s) 
    for the 2+3 peak (thick solid and dotted lines). 
    The model results are based on the 6 equations, generalization of
    Eqs.~(\ref{eqn.full.zonly}), for $u_{x,n}$ and $u_{z,n}$ with $n=2$, 3 and 5. 
    $X_n,Y_n = |u_{z,n,+} \pm u_{z,n,-}^*|$ are amplitudes of $X_n(t)$ and $Y_n(t)$.
    $Y_n$ becomes nonzero together with nonzero 
    $J_{\rm pseudo}$ above the critical sliding speed, for ``$2+3$'' excitations. 
    Hysteresis is not seen in the simulation. }
\end{figure}

%
{\it ``$\mathit{2+3=5}$'' excitation} - Consider now the main frictional peak labelled ``$2+3$'' 
in Fig.~\ref{fig.force.and.momentum}b, around $v\approx 570$ m/s. 
This corresponds, as shown by Fig.~\ref{fig.amplitudes.2+3}, to a mixed-resonance 
involving three modes, $n=2$, $3$, and $5$ simultaneously 
(see Fourier spectrum in Fig.~\ref{fig.force.and.momentum}c). 
The inset in Fig.~\ref{fig.amplitudes.2+3} shows that the cross-section snapshot 
of both tubes in this region of $v$, where frictional excitation is huge, is mostly 
elliptical, indicating a dominant $n=2$ deformation (see also the large $n=2$ 
peak in the Fourier spectrum of Fig.~\ref{fig.force.and.momentum}c), with 
a slight triangular ($n=3$) deformation. 
Our anharmonic continuum model nicely explains these results, the Umklapp terms 
%
%
(present in nanotubes but not in the string) playing a crucial role \cite{zhang_long}. 
In this region of $v$ we can truncate the full hierarchy of coupled equations to 6 equations 
involving the relevant Fourier modes $u_{x/z,n}$ with $n=2$, 3, 5 \cite{zhang_long}. 
%
%
To give a flavor of the full theory \cite{zhang_long}, we report here the equation of motion 
for mode $u_{z,5}$, obtained under the linear approximation $u_{x,5}\approx (i/5)u_{z,5}$ \cite{zhang_long},
\begin{eqnarray}
  \label{eqn.full.zonly}
  &&\ddot u_{z,5} = - \gamma \dot{u}_{z,5}
                    - Q u^*_{z,{5}} 
                    - (\omega_5^2 + \sum_{m=2,3,5} B_{m} |u_{z,m}|^2 ) u_{z,5} \cr
  &&                - A^{\rm dir} u_{z,{2}} u_{z,3} + A^{\rm umkl} u^*_{z,{2}} u^*_{z,3}
                    + 2F \cos(\Omega_w t),
\end{eqnarray}
where the last term represents the washboard forcing (due to inner tube sliding), 
while the $Q$-term is the previously discussed 
Umklapp term $-5 \to 5$. 
(Similar equations hold for the modes 
$n=2,3$, except
that the Q-term and the explicit forcing are missing.)
The physics of the $B_m$-terms is an effective {\em increase} of
the bare 
frequency $\omega_5$ due to the excitation of modes $m=2,3,5$,
via fourth-order non-linearities.
$A^{\rm dir}$ and $A^{\rm umkl}$ represent the effect of third-order
non-linearities, whereby a combined excitation of modes $2$ and $3$ can lead, by angular
momentum conservation, to a term influencing mode $5$ either in a direct way,
via $2+3=5$, or through Umklapp, via $-2-3=-5 \to 5$.
The stationary solutions of the nonlinear coupled equations for $u_{x/z,n=2,3,5}$, 
obtained numerically, reveal a rather simple structure, namely, 
$u_{x/z,5}(t) = u_{x/z,5,+} e^{i\Omega_w t} + u_{x/z,5,-} e^{-i\Omega_w t}$, 
$u_{x/z,2}(t) = u_{x/z,2,+} e^{i\Omega_2 t}$, and $u_{x/z,3}(t) = u_{x/z,3,+} e^{i\Omega_3 t}$, 
with $\Omega_2+\Omega_3=\Omega_w$: in words, mode $5$ oscillates with the driving washboard 
frequency $\Omega_w$, while modes $2$ and $3$ ``feel'' an indirect driving from mode $5$ and 
oscillate at frequencies $\Omega_2$ and $\Omega_3$ (renormalized by tube-tube coupling and
anharmonicity) in such a way as to realize a perfect ``resonance'' with the driving frequency
(see Fourier spectra Fig.~\ref{fig.force.and.momentum}c).

As simulations show, the amplitudes of $X_n=u_{z,n,+}+u^*_{z,n,-}$ and 
$Y_n=u_{z,n,+}-u^*_{z,n,-}$ (in qualitative agreement between model and simulation) 
are equal to each other for both $n=2$ and $n=3$, implying $u_{z,2/3,-}=0$. 
That means, in full agreement with the model, a full chirality for both modes, 
with waves travelling totally clockwise (or counter-clockwise, with equal probability). 
The $n=5$ $X$ and $Y$ amplitudes are, on the contrary, unequal and much smaller than $n=2$ and $3$. 
The continuum model actually predicts -- in striking agreement with the simulation -- that
the $n=5$ mode rotates {\it opposite} to the chirality of modes 2 and $3$ 
(i.e., counter-clockwise, $|u_{z,5,-}|>|u_{z,5,+}|$, for the case shown here),
%
%
an effect entirely due to {\em Umklapp} terms \cite{zhang_long}.
%
%

So far we have explained the two main features labelled A in Fig.~\ref{fig.force.and.momentum}b.
The features labelled B appear to represent a replica, where single outer tube $n=3,5$  
excitations replace the double tube ones $n=3,5$. For example, in the spectrum for 455 m/s 
(Fig.~\ref{fig.force.and.momentum}c), the $n=3s$ mode is at 1.14 THz, the outer tube value, 
different from the joint $n=3$ mode where both tubes vibrate together. 
Analogously, the $n=5s$ mode is excited at 720 m/s. Here, however, the friction is so small 
that the excited angular momentum is essentially invisible.
%
The two regions labelled C, finally, are irrelevant to the present context and will
be discussed elsewhere \cite{zhang_long}. 
%
%

In summary we have described a spontaneous breaking of chiral symmetry occurring
in nanoscale friction. Our very large speeds and termination-free conditions differ strongly 
from experimental conditions where nanotube sliding was observed 
so far.\cite{cumings.j:2000}.
Attempts at observing this chirality breaking could nonetheless be of considerable interest,
considering in addition the possibility to examine the effect of electronic frictional dragging\cite{persson_tosatti} 
an avenue which seems worth considering. Although the modes that play
a major role in electron-phonon coupling\cite{mauri}, and nanotube superconductivity
\cite{ferrier.m:2006} are different, breathing modes excitation was recently 
demonstrated by STM tips \cite{dekker}.
The physical understanding obtained for our microscopic nonlinear dynamical system 
might also serve as a prototype for phenomena in the field of nanomotors.
Examples of spontaneous chiral symmetry breaking that are relevant to the dynamics of
nanomotors have been described in rotaxanes \cite{hernandez.jv:2004}, where a
series of chemical reactions can give rise to a biased Brownian motion of a
small ring around a larger one in either direction. Our example represents a first
idealized case where the origin of nanoscale chiral symmetry breaking is strictly physical.

%
This research was partially supported by PRIN 2006022847, and by 
CNR/ESF/EUROCORES/FANAS/AFRI. We thank D. Ceresoli for an illuminating suggestion.
%
%

\begin{thebibliography}{10}

\bibitem{elliott.ja:1982}
J.~A. Elliott, Am. J. Phys. {\bf 50},  1148  (1982).

\bibitem{Guyon}
E. Guyon, J.-P. Hulin, L. Petit, and C.~D. Mitescu, {\em Physical
  hydrodynamics} (Oxford University Press, 2001).

\bibitem{lenz.p:2003}
P. Lenz {\em et al.}, Phys. Rev. Lett. {\bf 91},  108104  (2003).

\bibitem{brenner.dw:1990}
D.~W. Brenner, Phys. Rev. B {\bf 42},  9458  (1990).

\bibitem{kolmogorov.an:2005}
A.~N. Kolmogorov {\em et al.}, Phys. Rev. B {\bf 71},  235415  (2005).

\bibitem{zhang_ecoss}
X.-H. Zhang {\em et al.}, Surf. Sci. {\bf 601}, 3693  (2007).

\bibitem{soddemann.t:2003}
T. Soddemann {\em et al.}, Phys. Rev. E {\bf 68},  046702 (2003).

\bibitem{tangney.p:2006}
P. Tangney {\em et al.}, Phys. Rev. Lett. {\bf 97},  195901 (2006).

\bibitem{suzuura.h:2002}
H. Suzuura {\em et al.}, Phys. Rev. B {\bf 65},  235412  (2002).

\bibitem{zhang_long}
X.~H. Zhang {\em et al.}, in preparation.

\bibitem{cumings.j:2000}
J. Cumings {\em et al.}, Science {\bf 289},  602  (2000).

\bibitem{persson_tosatti}
B.~N.~J. Persson {\em et al.}, Phys. Rev. B {\bf 69},  235410  (2004).

\bibitem{mauri}
S. Piscanec {\em et al.}, Phys. Rev.  B {\bf 75},  035427  (2007).

\bibitem{ferrier.m:2006}
M. Ferrier {\em et al.} Phys. Rev. B {\bf 74},  241402(R)  (2006).

\bibitem{dekker}
B.~J. LeRoy {\em et al.}, Nature {\bf 432},  371 (2004).

\bibitem{hernandez.jv:2004}
J.~V. Hern\'{a}ndez {\em et al.}, Science {\bf 306},  1532 (2004).

\end{thebibliography}

%
%
\end{document}